\documentclass[onecolumn,draftclsnofoot,12pt]{IEEEtran}                                
\usepackage{graphicx}
\usepackage{amsmath,amsthm} 
\usepackage{color}                      
\usepackage{algorithm,algorithmic}
\usepackage{graphics} 
\usepackage{epstopdf} 
\usepackage{amsmath} 
\usepackage{amssymb}  
\graphicspath{{../pdf/}{../jpeg/}}
\DeclareGraphicsExtensions{.pdf,.jpeg,.png}
\newtheorem{lem}{Lemma}
\def\x{\mathbf{x}}
\def\c{\mathbf{c}}

\title{On Socially Optimal Traffic Flow in the Presence of Random Users
}
\author{ \centering Anant Chopra, Deepak S. Kalhan, Amrit S. Bedi, \\Abhishek K. Gupta, Ketan Rajawat  \\

\thanks{The authors are with the Department of Electrical Engineering, Indian Institute of Technology Kanpur, Kanpur 208016, India.}
} 

\newcommand{\abhiadd}[1]{{\color{blue}#1}}
\renewcommand{\abhiadd}[1]{#1}

\providecommand{\c}[1]{\mathcal{#1}}

\providecommand{\c}[1]{\mathcal{#1}}

\newcommand{\Ex}[1]{\mathbb{E}\left[#1\right]}

\begin{document}

\maketitle
\thispagestyle{empty}
\pagestyle{empty}


\begin{abstract}

Traffic assignment is an integral part of urban city planning. Roads and freeways are constructed to cater to the expected demands of the commuters between different origin-destination pairs with the overall objective of  minimising the  travel cost. As compared to static traffic assignment problems where the traffic network is fixed over time, a dynamic traffic network is more realistic where the network's cost parameters change over time due to the presence of random congestion. 
In this paper, we consider a stochastic version of the traffic assignment problem where the central planner is interested in finding an optimal social flow in the presence of random users. These users are random and cannot be controlled by any central directives. We propose a Frank-Wolfe algorithm based stochastic algorithm to determine the socially optimal flow for the stochastic setting in an online manner. Further, simulation results corroborate the efficacy of the proposed algorithm.

\end{abstract}

\section{Introduction}\label{intro}

Traffic congestion is considered to be one of the most important problems of any urban city. A large number of possible routes to reach a particular destination makes it difficult for the user to choose the optimal path. Optimality of a path itself depends on the user's perspective. Generally, a user would choose  a path familiar to him/her. However, there may be situations when a user has to divert from his regular known path and is unable to choose the best path from the existing options. A user may also rely completely on the  navigation system which can guide the user to his/her destination. The navigation strategy may be based on both, the individual or the social perspective. The individual perspective to decide a route would be to choose the least costly available path. This strategy may lead to an equilibrium which is the Nash equilibrium of corresponding congestion game and can be formed as an optimization problem using Wardrop's equilibrium conditions \cite{wardrop1952some}. However, this strategy is not optimal from the perspective of the city government or the city planner (e.g. department of transportation) \cite{LEBLANC1975309}. The city planner would want the total travel cost in the city to be minimized, which is popularly also known as the social optimal situation.

The earlier traffic assignment problems have been primarily considered in two ways: The \textit{static} traffic assignment problem and the \textit{dynamic} traffic assignment problem \cite{ksaw_lit_rev}. In static traffic assignment problem, the demands for various origin-destination pairs are considered to be constant in time. Such scenarios represent networks where the traffic would inherently be constant, like in railway networks. However, if demands are assumed as functions of time, the problem is known as a dynamic traffic assignment problem. These time varying demand situations arise during emergency evacuation in cases of natural hazards \cite{ksaw_lit_rev}.

Static traffic assignment doesn't consider variation in time which makes it difficult to tackle traffic issues \cite{Liu_2005}, but it is still preferred because of its simplicity which makes it useful to determine optimal solutions for large networks in  less time \cite{Michel_2013}. On the other hand, dynamic traffic assignment estimates the traffic conditions based on the historical and real time data analysis. This utilization of available data provides support in critical conditions and gives better results than the static traffic assignment \cite{Han_2010}. The problem of traffic assignment belongs to the class of constrained convex optimization problem \cite{LEBLANC1975309}.  Need for the dynamic traffic assignment can also arise in the scenario where the demands are fixed but the traffic network itself changes due to uncontrollable random congestion in the network. This random congestion can be due to users not following centralized directives, or an unexpected change in road conditions and may lead to a very different optimal solution. The past work has not considered such a dynamic network which is the main focus of this work.

To solve the problem in an iterative manner, one promising technique is to use the Frank-Wolfe Algorithm to determine the optimal flows \cite{LEBLANC1975309}. We consider stochastic optimization framework because stochastic algorithms operate on stochastic estimates of objective function gradients, decreasing the computational complexity \cite{mokhtari2018stochastic}. In literature, projected stochastic gradient (SGD) \cite{robb} and stochastic variants of Frank-Wolfe algorithm \cite{mokhtari2018stochastic} have been used to minimize a stochastic convex function subject to a convex constraint. In SGD algorithm,  iterations are performed by descending through the negative direction of stochastic gradient with a proper step size and projecting the resulted point onto the feasible set $\cite{mokhtari2018stochastic}$. 


In this paper, we consider a stochastic version of the traffic assignment problem which takes care of the inherent randomness present on each link of the traffic network. This corresponds to the practical situation where there are always some travelers who either do not follow centeralized directives such as  recommendation of navigation systems to attain the optimal solution or travelers who are unaccounted for in the system database like rickshaws and autos in a typical suburban city for example an indian city. We propose a traffic assignment algorithm based on the stochastic Frank-Wolfe algorithm \cite{mokhtari2018stochastic}  to solve the problem in an online manner by considering the current samples of random additional congestion. We further state that the assumptions required to apply the algorithm of \cite{mokhtari2018stochastic} are satisfied. The simulation results describe the efficacy of the proposed algorithm.

\section{Problem Formulation}
\newcommand{\trafficmat}{\mathbf{D}}
\newcommand{\mnflow}{$(m,n)-$}
\newcommand{\ijflow}{$(i,j)-$}
\newcommand{\flow}{f}
\newcommand{\flowfunction}{g}
\newcommand{\detflow}{x}
\newcommand{\randflow}{z}

In this paper, we consider a traffic network in an area as a directed graph $\mathcal{G} = (\mathcal{N}, \mathcal{E})$ where $\mathcal{N}$ is the ordered list of nodes and $\mathcal{E}$ is the ordered list of edges or links on which the traffic flows. Each node of the network may serve as an origin or destination or both for the traffic flow. For simplicity, we assume that multiple edges in the same direction between any two nodes do not exist. For an edge $e \in \mathcal{E}$, let $\flow_e$ denotes the flow on  edge $e$ and $\mathbf{\flow}=[\flow_1,\flow_2,\cdots, \flow_e, \cdots \flow_{|\mathcal{E}|}]^T$ be the flow vector of length $\left| \mathcal{E} \right|$. Let $\mathcal{P}_{mn}$ denotes the set of all open loop paths from the $m^{th}$ node to the $n^{th}$ node. Let \mnflow flow mean the flow wishing to move from node $m$ to node $n$. For a path $p \in \{ \mathcal{P}_{mn}\}$, let $\flow_p$ denotes the \mnflow flow on this path $p$. 
Let $\trafficmat$ be a $\left| \mathcal{N} \right| \times \left| \mathcal{N} \right| $ matrix whose $(i,j)^{th}$ element, $D_{ij}$, denotes the mean flow of commuters wishing to move from the node $i$ to the node $j$. Now, the total \ijflow traffic flow over all possible paths between two nodes $i$ and $j$ should be equal to the total demand between these nodes {\em i.e.}
\begin{equation}\label{const_1}
D_{ij} = \sum_{p \in \mathcal{P}_{ij}} \flow_p, \hspace{0.3cm} \forall \hspace{0.1cm} i,j = 1,2,\ldots, \left| \mathcal{N} \right|.
\end{equation}
Also, for any edge, the total flow in it is equal to the total flow over all paths which consists of this edge {\em i.e.}
\begin{equation}\label{const_2}
\flow_e = \sum_{p \ni e} \flow_p,  \hspace{0.3cm} \forall \hspace{0.1cm} e \in \mathcal{E}.
\end{equation}
Additionally, it is important to note that all flows are non-negative
\begin{equation}\label{const_3}
\flow_p \geq 0 \hspace{0.3cm} \forall \hspace{0.1cm} p.
\end{equation}
There is an inherent traversal cost associated with each link which may be a function of travel time, distance, congestion, tension, fuel or a combination of these \cite{LEBLANC1975309}. 
For any commuter, let $C_e$ denote the cost to traverse link $e$:
\begin{align*}
C_e=c_e(\flow_e)
\end{align*}
where $c_e$ denotes the cost function of link $e$ and $\flow_e$ denotes the total flow in this link. In particular, 
\begin{equation}\label{cost_function}
c_e(\flow_e) = a_{e} + b_{e} \cdot {\flow_{e}}^{4}, \hspace{0.5cm} \forall \hspace{0.1cm} e \in \mathcal{E}
\end{equation}
where $a_{e}$ and $b_{e}$ are fixed parameters corresponding to the link $e\in\mathcal{E}$. We can see that the cost function in \eqref{cost_function} for each link is a convex function. It is remarked that although we have considered a particular type of cost function for simplicity, but the analysis and the proposed algorithm can be extended for other convex cost functions. Let $\mathbf{c}=[C_1,C_2,\cdots, C_e, \cdots C_{|\mathcal{E}|}]^T$ be the cost vector of length $\left| \mathcal{E} \right|$.

In general, the objective function for a traffic assignment problem depends on the various strategies that travelers choose for themselves. In this paper, we consider a social optimal model \cite{wardrop1952some}, in which there is a central authority such as city planning or traffic department which can direct the flow of all commuters into various paths with the goal to minimize the total cost of all commuters. Mathematically, this corresponds to the following objective function
\begin{align}\label{social_opt_objective}
\flowfunction(\mathbf{\flow})=\sum_{\forall \hspace{0.05cm} e\in \mathcal{E}} \flow_{e} \cdot c_{e}(\flow_{e})
\end{align}
and the goal is to choose an optimal $\mathbf{\flow}^\star$ flow which  minimizes the cost in \eqref{social_opt_objective} subject to constraints in \eqref{const_1}, \eqref{const_2}, and \eqref{const_3}. 

This model works in the deterministic scenario where each commuter follows the central directive. In reality, traffic flows are not so deterministic and there is always some inherent randomness to them. This can be due to the unexpected commuters or commuters not adhering to directive or some other unexpected reasons. In this paper, we model the unexpected commuters via including randomness to each link flow. This additional flow $z_e$  changes with time  to account for the stochastic and dynamic nature of the traffic. The optimal flow vector $\mathbf{\flow}$ in a stochastic environment would be different than the ones obtained using objective functions \eqref{social_opt_objective}. In this case, it is of interest to talk about a best 'average' flow. We aim to achieve this \textit{stochastic social optimal} solution.

Let $\mathbf{z}$ be a vector of length $\left| \mathcal{E} \right|$ whose element, $z_{e}$, is a random variable which denotes the uncertain flow on link $e$. The elements of vector $\mathbf{z}$  follows an unknown probability distribution. Without loss of generality, we assume that $z_e$ has zero mean i.e.
\begin{equation}\label{zero_mean}
    \mathbb{E}[z_{e}] = 0, \hspace{0.3cm}\forall \hspace{0.1cm}e \in \mathcal{E}.
\end{equation}
Note that  a non-zero mean implies that on average the mean number of travelers use that link and if this was the case, then the mean traffic can be included as a part of the deterministic flow itself.

Now the flow in each link $e$ can be written as the summation of the deterministic flow $\detflow_e$ and the random flow $\randflow_e$:
\begin{align}
\flow_e&=\detflow_e+\randflow_e.
\end{align}
Note that deterministic flow $\detflow_e$ is also the average flow in the link as
\begin{align}
\mathbb{E}[\flow_e]&=\mathbb{E}[\detflow_e]+\mathbb{E}[\randflow_e]=\detflow_e.
\end{align}
%
%
Since, the central authority can only control the deterministic or the average flow $\detflow_e$ to optimize the cost. The problem of the social optimum flow in a stochastic environment with fixed average demands can be formulated as

\begin{align}\label{objective_func}
    \min_{x_e} & \hspace{0.2cm}\mathbb{E} \left [\sum_{\forall e \in \mathcal{E}} (x_{e}+z_{e}) \cdot c_{e}(x_{e}+z_{e}) \right]\\
\text{subject to } & D_{ij} = \sum_{p \in \mathcal{P}_{ij}} \detflow_p \hspace{0.3cm} \forall \hspace{0.1cm} i,j = 1,2,\ldots, \left| \mathcal{N} \right|\label{cons_1}\\
&
x_p \geq 0 \hspace{0.3cm} \forall \hspace{0.1cm} p\label{cons_2}
\\
&x_e = \sum_{p \ni e} x_p  \hspace{0.3cm} \forall \hspace{0.1cm} e \in \mathcal{E}\label{cons_3}
\end{align}
where expectation in \eqref{objective_func} is taken with respect to $\textbf{\randflow}$. Note that the demand constraints are only put on the average flow as the demand matrix is itself an average demand over time.

\section{Proposed Online Algorithm}\label{model_descrip}
We have a stochastic constrained optimization problem in \eqref{objective_func} and seek to solve it in an online manner \cite{hazan2016introduction}. From online we mean that the decisions are taken`on the fly' when the random variables are realized. It is emphasized that since the distributions of the random variables in \eqref{objective_func} is not known, it is not possible to solve the problem in closed form. Hence, we are interested in developing an online solution to the problem in \eqref{objective_func} which works when the random variables are realized iteratively one by one. In literature, there are different online approached which can be utilized to solve the problem such as dual ascent methods \cite{Rib_erg}, saddle point methods \cite{M_trad_regret}. But in these algorithms, the constraints are not satisfied at every step of the algorithm which is crucial to the traffic assignment problem considered here. 
\abhiadd{Therefore, we propose a stochastic version of the traffic assignment algorithm \cite{LEBLANC1975309}. This algorithm is based on the stochastic Frank-Wolfe algorithm recently proposed in the literature \cite{mokhtari2018stochastic}.}

\newcommand{\dd}{\mathrm{d}}
\newcommand{\expect}[1]{\mathbb{E}\left[#1\right]}
\abhiadd{
For implementation of the proposed stochastic Frank-Wolfe traffic assignment algorithm (SFWTA), we need to compute the gradient of the stochastic and expected objective function. Let the stochastic objective function be
\[\tilde{F}(\mathbf{x},\mathbf{z})=\sum_{e\in\mathcal{E}} (x_e+z_e)c_e(x_e+z_e)\]
then the gradient is given as
\begin{align}
\nabla\tilde{F}(\mathbf{x},\mathbf{z})&= \{\nabla\tilde{F_e}(\mathbf{x},\mathbf{z})\}
\end{align}
where 
\[\nabla\tilde{F_e}(\mathbf{x},\mathbf{z})=\frac{\dd}{\dd x_e}\tilde{F}(x_e,z_e)\]
Let dependence of $z_e$ on $x_e$ is modelled as {\em i.e. $z_e=z(x_e)$}. Then
\begin{align}
\nabla\tilde{F_e}(\mathbf{x},\mathbf{z})&=\frac{d}{dx_e}a_e (x_e+z(x_e))+ b_e (x_e+z_e)^5\nonumber\\
&=a_e (1+z'(x_e))+ b_e 5(x_e+z_e)^4 (1+z'(x_e)).\nonumber
\end{align}

Let $F(\mathbf{x})$ denote the average objective function {\em i.e.}
\begin{align}
F(\mathbf{x})=\expect{\tilde{F}(\mathbf{x},\mathbf{z})}=\expect{\sum_{e\in\mathcal{E}} (x_e+z_e)c_e(x_e+z_e)}
\end{align}
The average gradient is 
\begin{align}
\nabla F(\mathbf{x})&=\{\nabla F_e(\mathbf{x})\} \text{ where}\nonumber\\
\nabla F_e(\mathbf{x})&=\expect{\nabla\tilde{F_e}(\mathbf{x},\mathbf{z})}\nonumber\\
&=a_e (1+\expect{z'(x_e)})+ 5 b_e \expect{(x_e+z_e)^4 (1+z'(x_e))}.\nonumber
\end{align}

\textbf{Special cases}:
\begin{enumerate}
\item
Case 1: if $z_e=x_e u_e$ where $u_e$ is independent uniformly distributed between $[-1,1]$, then 
\begin{align}
\nabla\tilde{F_e}(\mathbf{x},\mathbf{u})\nonumber&=
a_e (1+u_e)+ 5 b_e x_e^4 (1+u_e)^5\nonumber\\
\nabla F_e(\mathbf{x})&=\expect{\nabla\tilde{F_e}(\mathbf{x},\mathbf{u})}
\nonumber=
a_e + 5 b_e x_e^4 \frac{16}{3}\nonumber.
\end{align}
\item Case 2: Let $z_e$ is independent to $x_e$, then
\begin{align}
\nabla\tilde{F_e}(\mathbf{x},\mathbf{z})&=a_e + 5b_e (x_e+z_e)^4,\nonumber\\
\nabla F_e(\mathbf{x})=\expect{\nabla\tilde{F_e}(\mathbf{x},\mathbf{z})}&=a_e + 5b_e\sum_{i=0}^{{4}} \binom{4}{i}x_e^i\expect{z_e^{4-i}}.
\nonumber
\end{align}
\end{enumerate}

Hence, the biased-gradient-estimate update step at iteration $t$ is given as
\begin{align*}
\mathbf{d}_t&=(1-\rho_t)\mathbf{d}_{t-1}+\rho_t\nabla \tilde{F}(\mathbf{x},\mathbf{z})
\end{align*}
where $\mathbf{d}_t$ is the biased-gradient-estimate at iteration $t$. The descent direction $\mathbf{y}$ is found using \cite[Theorem]{LEBLANC1975309} which is given as the step 6 in \textbf{Algorithm 1}. The complete proposed SFWTA algorithm is stated as \textbf{Algorithm 1}.
}

In order to utilize the algorithm of \cite{mokhtari2018stochastic}, it is important to satisfy the assumptions made by \cite{mokhtari2018stochastic} for the problem of interest here. Hence, next we present the assumptions required and their proof of being satisfied for the traffic assignment considered in this paper.
\begin{algorithm} 
\caption{Stochastic Traffic Assignment Algorithm}\label{algo}
\begin{algorithmic} [1]

\STATE \textbf{Require} step sizes $\rho_t > 0$ and $\gamma_t > 0$ where t is the iteration number.\\
Choose $\rho_t$ and $\gamma_t$ such that $(i) \sum_{t=0}^{\infty}\rho_t=\infty, \hspace{0.1cm}(ii) \sum_{t=0}^{\infty}\rho_t^2<\infty,
\hspace{0.1cm}(iii) \sum_{t=0}^{\infty}\gamma_t=\infty,  \hspace{0.1cm}(iv) \sum_{t=0}^{\infty}(\gamma_t^2/\rho_t)<\infty           $

\STATE \textbf{Initialize} $t=0$.\\
\textbf{Initialize} the deterministic flow vector $\mathbf{x}$.\\
\textit{One initialization method is to use the Dijkstra's algorithm to assign the entire flows of each origin-destination pair to the least cost path between the nodes.\\}
\textbf{Initialize} the cost vector $\mathbf{c}(\mathbf{f})$ as a zero vector.

\STATE \label{sample}\textbf{Sample} the random-flow vector $\mathbf{z}$.\\
Sample the elements of $\mathbf{z}$ using properties (\ref{zero_mean}) and (\ref{random_prop1}).

\STATE \label{random_values_added}\textbf{Randomize} $\mathbf{f}$:
$$\mathbf{f} = \mathbf{x} + \mathbf{z}.$$

\STATE \label{cost_step}\textbf{Update} the cost vector $\mathbf{c}(\mathbf{f})$.\\
Compute vector $\mathbf{c'}$ whose element, $c'_{e} =
\nabla \tilde{F}(x_e,z_e).$

$\mathbf{c} = (1 - \rho_t)\mathbf{c} + (\rho_t)\mathbf{c'};$

\STATE \textbf{Assign} all demands to the shortest paths using link costs as obtained in step (\ref{cost_step}) above and compute the corresponding flow vector $\mathbf{y}$.

\STATE \textbf{Update} the flow vector $\mathbf{x} = (1-\gamma_{t+1})\mathbf{x} + (\gamma_{t+1})\mathbf{y}$

\STATE \textbf{Test} the stopping criteria.\\
If test fails, update $t=t+1$. Go to step (\ref{sample}). \\
Else exit.\\
\textit{One exit criteria is to determine the maximum relative change in the elements of vector $\mathbf{x}$ and test it against a pre-determined threshold value.}

\end{algorithmic}
\end{algorithm}
\subsection{Assumptions Required}
For the sake of completeness, we recite the assumptions mentioned in \cite{mokhtari2018stochastic} and show that they hold for our case.
\begin{itemize}
\item {\textit{Assumption 1}}: \textit{The convex set $\mathcal{C}$ is bounded with diameter $K_\mathrm{max}$, i.e., for all$\hspace{0.1cm} \mathbf{x}, \mathbf{y}\hspace{0.1cm} \in \hspace{0.1cm}$C,  we can write}
\begin{equation}
||\mathbf{x}-\mathbf{y}||\leq K_\text{max}
\end{equation}

\abhiadd{
The set $\mathcal{C}$ in our case is denoted by the feasible convex set of constraints in \eqref{cons_1}-\eqref{cons_3}.
For any edge $e$, let $S_e$ denote the set of all source-destination pair containing atleast one path  which contains this edge {\em i.e.} \[S_e=\{(m,n): \exists  p\in \mathcal{P}_{m,n} \text{ such that }p \ni e\}.\]
Let $K_e= \sum\limits_{(m,n)\in S_e} D_{m,n}$. 
From constraints, it is evident that 
$x_e\le K_e$. Now, if $\mathbf{x}\in\mathcal{C}$ then \[||\mathbf{x}||=\left[\sum_{e\in\mathcal{E}}\detflow_e^2\right]^{1/2}\le \left[\sum_{e\in\mathcal{E}}K_e^2\right]^{1/2}.\]
Now, 
\begin{align*}
||\mathbf{x}-\mathbf{y}||&\le  ||\mathbf{x}||+||\mathbf{y}||\le 2\left[\sum_{e\in\mathcal{E}}K_e^2\right]^{1/2}.
\end{align*}
Hence, the optimization problem $\mathcal{P}$ satisfies Assumption 1 with 
\begin{align*}
K_{\max}&=2\left[\sum_{e\in\mathcal{E}}{\left[\sum\limits_{(m,n)\in S_e} D_{m,n}\right]}^2\right]^{1/2}\\
&\le |\mathcal{E}|^{1/2}\max_e|S_e|\max_{ij}D_{ij}.
\end{align*}
This proves the assumption 1.
}
 
\item {\textit{Assumption 2}}: \textit{The expected function $F$ is convex. Moreover, its gradients $\nabla F$ are L-Lipschitz
continuous over the set $\mathcal{C}$, i.e., for all $\mathbf{x}, \mathbf{y}\hspace{0.1cm} \in\hspace{0.1cm}$ $\mathcal{C}$
}\begin{equation}\label{assmp_2}
||\nabla F(\mathbf{x}) - \nabla F(\mathbf{y})|| \leq L||\mathbf{x}-\mathbf{y}||
\end{equation}
Recall that 
\begin{equation}
    F(\mathbf{x}) = \mathbb{E} \left [\sum_{\forall e \in \mathcal{E}} (x_{e}+z_{e}) \cdot c_{e}(x_{e}+z_{e}) \right]
\end{equation}
where $c_{e}$ is defined in equation (\ref{cost_function}) and $\hspace{0.1cm}x_{e},\hspace{0.1cm}z_{e}$ are ${e}^{\text{th}}$ elements of vectors $\mathbf{x},\mathbf{z}$ respectively. Next, utilizing the cost function defined in \eqref{cost_function} and vector form of $a_e$  and $b_e$, the gradient vector $\nabla F(\mathbf{x})$ is written as
\begin{equation}\label{gradfx}
 \nabla   F(\mathbf{x}) = \hspace{0.2cm}\mathbb{E} \left[\mathbf{a}+ 5\mathbf{b}\circ\tilde{\mathbf{x}} \right], 
\end{equation}
where $\tilde{\mathbf{x}} = \left(\mathbf{x}+\mathbf{z}\right)^{\circ 4}$. Here, we represents the $\textit{Hadamard Product}$ using symbol  '$\circ$' and hence we write  
$$\left(\mathbf{x}+\mathbf{z}\right)^{\circ 4}= \left(\mathbf{x}+\mathbf{z}\right)
{\circ } \left(\mathbf{x}+\mathbf{z}\right)\circ\left(\mathbf{x}+\mathbf{z}\right)\circ(\mathbf{x}+\mathbf{z}).$$ From equation \eqref{gradfx}, we can write
\begin{equation}\nonumber
[\nabla F(\mathbf{x}) - \nabla F(\mathbf{y})] = \mathbb{E}\left[5 \mathbf{b}\circ ( (\mathbf{x}+\mathbf{z})^{\circ 4} - (\mathbf{y}+\mathbf{z})^ {\circ 4}) \right].
\end{equation}
Further, each element of $[\nabla F(\mathbf{x}) - \nabla F(\mathbf{y})]$ can be simplified as
\begin{multline}\label{lipsch}
\mathbb{E}\left[5 {b}_{e} ( ({x}_{e}+{z}_{e})^{\circ 4} - ({y}_{e}+{z}_{e})^ {\circ 4}) \right] \\= \mathbb{E} [5b_{e}(x_{e}- y_{e})
(x_{e}+y_{e}+2z_{e})((x_{e}+z_{e})^2 + (y_{e}+z_{e})^2)].
\end{multline}
In general, equation \eqref{lipsch} can be written as
\begin{equation}
[\nabla F(\mathbf{x}) - \nabla F(\mathbf{y})] = \mathbf{g} \circ (\mathbf{x}-\mathbf{y}),
\end{equation}
where $\mathbf{g}$ is a vector with terms of the form $\mathbb{E}[5b_{e}(x_{e}+y_{e}+2z_{e})((x_{e}+z_{e})^2 + (y_{e}+z_{e})^2)]$.
If we take norm both sides and utilize  the result \cite[Theorem 2.5]{horn1990analog}, we get
\begin{equation}
||\nabla F(\mathbf{x}) - \nabla F(\mathbf{y})|| \leq L||\mathbf{x}-\mathbf{y}||
\end{equation}
where $L$ is the upper bound on the norm of $\mathbf{g}$, which can be calculated from result in Assumption 1 and from the fact that $|z_e|\leq x_e$.
Thus, Assumption 2 holds for our case.
\item {\textit{Assumption 3}}:\textit{ The variance of the unbiased stochastic gradients $\nabla \tilde{F}$($\mathbf{x}$, $\mathbf{z}$) is bounded above by $\sigma^2$.} To prove this, note that
\begin{equation}
  \tilde{F}(\mathbf{x}, \mathbf{z}) =   \hspace{0.2cm} \left (\sum_{\forall \hspace{0.1cm} e \in \mathcal{E}} (x_{e}+z_{e}) \cdot c_{e}(x_{e}+z_{e}) \right),
\end{equation}
which implies that
\begin{equation}\label{un_st_grad}
\nabla\tilde{F}(\mathbf{x}, \mathbf{z}) = \hspace{0.1cm} \left[\mathbf{a}+5 \mathbf{b}\circ(\mathbf{x}+\mathbf{z})^{\circ 4}\right].
\end{equation}
From \eqref{gradfx} we have,
\begin{equation}\label{st_gr}
 [\nabla   F(\mathbf{x})] = \hspace{0.2cm}\mathbb{E} \left[\mathbf{a}+ 5\mathbf{b}\circ({\mathbf{x}}+\mathbf{z})^{\circ 4} \right].
\end{equation}
It holds that $\mathbb{E}[\mathbf{a}]$ = $\mathbf{a}$ and $\mathbb{E}[\mathbf{b}]$ = $\mathbf{b}$ as elements of $\mathbf{a}$ and $\mathbf{b}$ are path factors which are deterministic. Using (\ref{un_st_grad}) and (\ref{st_gr}), we write
\begin{equation}
\nabla\tilde{F}(\mathbf{x}, \mathbf{z})-\nabla   F(\mathbf{x}) = 5 \mathbf{b}\circ\left((\mathbf{x}+\mathbf{z})^{\circ 4} -  \mathbb{E} \left[({\mathbf{x}}+\mathbf{z})^{\circ 4}\right]\right).
\end{equation}
If we take norm both sides and utilize \cite[Theorem 2.5]{horn1990analog},
\begin{align}
\mathbb{E}&||\nabla\tilde{F}(\mathbf{x}, \mathbf{z})-\nabla   F(\mathbf{x})||^2\nonumber \\
 &=\mathbb{E} ||5 \mathbf{b}\circ\left((\mathbf{x}+\mathbf{z})^{\circ 4} -  \mathbb{E} \left[({\mathbf{x}}+\mathbf{z})^{\circ 4}\right]\right)||^2.\\&\leq ||5 \mathbf{b}||^2\hspace{0.2cm}\mathbb{E}||(\mathbf{x}+\mathbf{z})^{\circ 4} -  \mathbb{E} \left[({\mathbf{x}}+\mathbf{z})^{\circ 4}\right]||^2.
\end{align}

In order to prove the bounded variance, considering the randomness at each edge $z_e$ to be independent of other edges, it is sufficient to show that variance of random variable $(f_e)^4=(x_e+z_e)^4$ is finite.
\abhiadd{We will prove this for the special case where  $z_e=x_eu_e \sim U[-x_e, x_e]$ with mean $\Ex{z_e}=0$ and variance $\Ex{(z_e-\Ex{z_e})^2}=x_e/6$.}  This implies that $f_e=x_e(1+u_e)\sim U[0,2x_e]$ with mean $x_e$ and variance \abhiadd{$x_e^2/6$}.
The moment generating function of $f_e$ is 
\begin{equation}
M(t) = (e^{2x_{e}t}-1)/2x_{e}t
\end{equation}
 Using this, the $r^\text{th}$ moment of $f_e$ is 
 \begin{equation}
 \mathbb{E}[{(f_e)^r}]=\frac{(2x_e)^r}{r+1}
 \end{equation}
 
Now,
\begin{align}
&\mathbb{E}\left[(f_e)^4 - \mathbb{E}[(f_e)^4]\right]^2 = \mathbb{E}\left[(f_e)^4\right]^2 - \{\mathbb{E}[(f_e)^4]\}^2\nonumber\\
&= \frac{(2x_e)^8}{9} - \left[\frac{(2x_e)^4}{5}\right]^2=\frac{16}{225}(2x_e)^8
\end{align}
 Since, $x_e$ is always bounded, the variance of $(f_e)^4$ will be bounded. From the bounded variance of $(f_e)^4$, we conclude that 
\begin{equation}
\mathbb{E}|| \nabla\tilde{F}(\mathbf{x}, \mathbf{z})-\nabla   F(\mathbf{x})||^2 \leq S_{\max}^2
\end{equation}
where $S_{\max}$ is a finite number.
\end{itemize}

Next, as we have proved that all the assumptions made in \cite{mokhtari2018stochastic} holds for the problem in this paper, we restate the convergence result from \cite{mokhtari2018stochastic} as follows. 
\begin{lem}
With all the Assumptions 1-3 satisfied, for the iterates $\x_t$ and $\c_t$ produced by the proposed Algorithm \ref{algo}, it holds that

\begin{align}
\mathbb{E}[||\nabla F(\mathbf{x}_t) - \mathbf{c}_t||^2\hspace{0.01cm} |\hspace{0.01cm}\mathcal{F}_t] &\leq \left(1-\frac{\rho_t}{2}\right)||\nabla F(\mathbf{x}_{t-1})-\mathbf{c}_{t-1} ||^2\nonumber\\
&\ \ \ \ + {\rho_t}^2{S_\textit{max}^2} + \frac{2 L^2{K_\text{max}^2}{\gamma_t^2}}{\rho_t},
\end{align}  
\end{lem}
\noindent where $\mathbf{x}_t$ \& $\mathbf{c}_t$ denote the vectors $\mathbf{x}$ \& $\mathbf{c}$ at the $t^{th}$ iteration and $\mathcal{F}_t$ represents the sigma algebra associated with all sources of randomness up to $t^\text{th}$ iteration.
From Lemma 1, it is clear that with every iteration, there is a decrease in the squared error of gradient approximation, provided the term ${\rho_t}^2{\sigma}^2 + \frac{2 L^2{K_\text{max}^2}{\gamma_t^2}}{\rho_t}$ is negligible and adhere to the step size requirements mentioned in Algorithm 1. 


\section{Examples and Numerical Results}\label{sim_res}
In this section, we provide numerical results for the proposed algorithm in terms of performance and the convergence of the algorithm. For this numerical evaluation, we consider a traffic network consisting of 4 nodes and single demand between $\mathsf{A}$ and $\mathsf{D}$ (refer Fig. 1). We have assumed that the unity flow of commuters wish to move from node $\mathsf{A}$ (node 1) to node $\mathsf{D}$ (node 4), i.e.,
$$\mathbf{D}=\begin{bmatrix}
0 & 0 & 0 & 1\\
0 & 0 & 0 & 0\\
0 & 0 & 0 & 0\\
0 & 0 & 0 & 0
\end{bmatrix}.$$

Let $\mathbf{a}$ and  $\mathbf{b}$ be the network parameter vectors whose elements, $a_e$ and $b_e$, denote the travel cost parameters associated with link $e$. We assume the network parameters to be:
\begin{figure}[t!]
\centering
\includegraphics[width= 3 in]{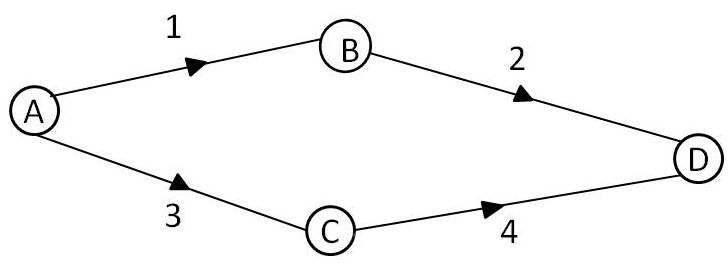}\label{node}
\caption{A illustration showing the example network (4 nodes \& 4 links) considered.}
\end{figure}

\begin{figure}[ht!]
\begin{center}
\includegraphics[width=0.6\linewidth]{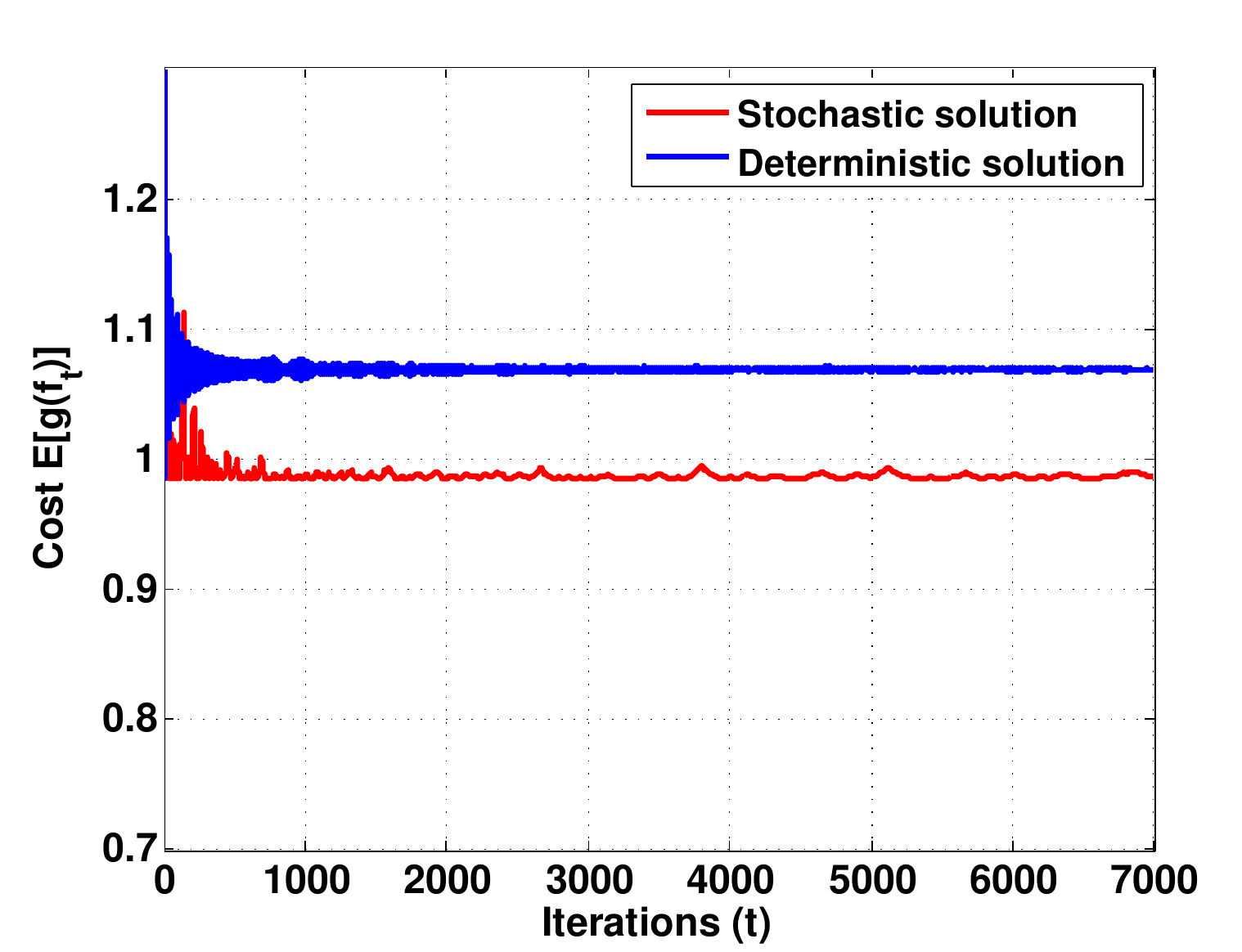}
\caption{Running mean of the costs of the stochastic \& deterministic solution strategies in stochastic environment with $\beta=1$.}\label{fig:numres1}
\end{center}
\end{figure}

\begin{equation*}
\mathbf{a} = \begin{bmatrix}
0.3\\
0.3\\
0.5\\
0.5\\
\end{bmatrix}
\hspace{0.5cm} \text{and } \hspace{0.5cm}
\mathbf{b} = \begin{bmatrix}
0.6\\
0.6\\
0.1\\
0.1\\
\end{bmatrix}.
\end{equation*}


The random traffic on link $e$ is assumed to have a uniform probability distribution between $-\beta x_{e}$ to $\beta x_{e}$ with the understanding that that  the number of random travelers on a link would not be more than the deterministic flow on that link at any point of time. In other words, we have
\begin{equation}\label{random_prop1}
\left| z_e \right| \leq \beta x_{e}.
\end{equation}
Here, $\beta$ is the spread parameter and $0\le\beta\le 1$. Note that $\beta=0$ denotes deterministic environment. Therefore, the total flow on a link $e$ can be represented as
\[\flow_e=x_{e} + z_{e}=x_e(1+\beta u_e)\]
where $u_e$ is uniform distributed random variable $\sim \mathsf{U}[-1,1]$ independent to $x_e$.

\begin{figure}[t!]
\begin{center}
\includegraphics[width=0.6\linewidth]{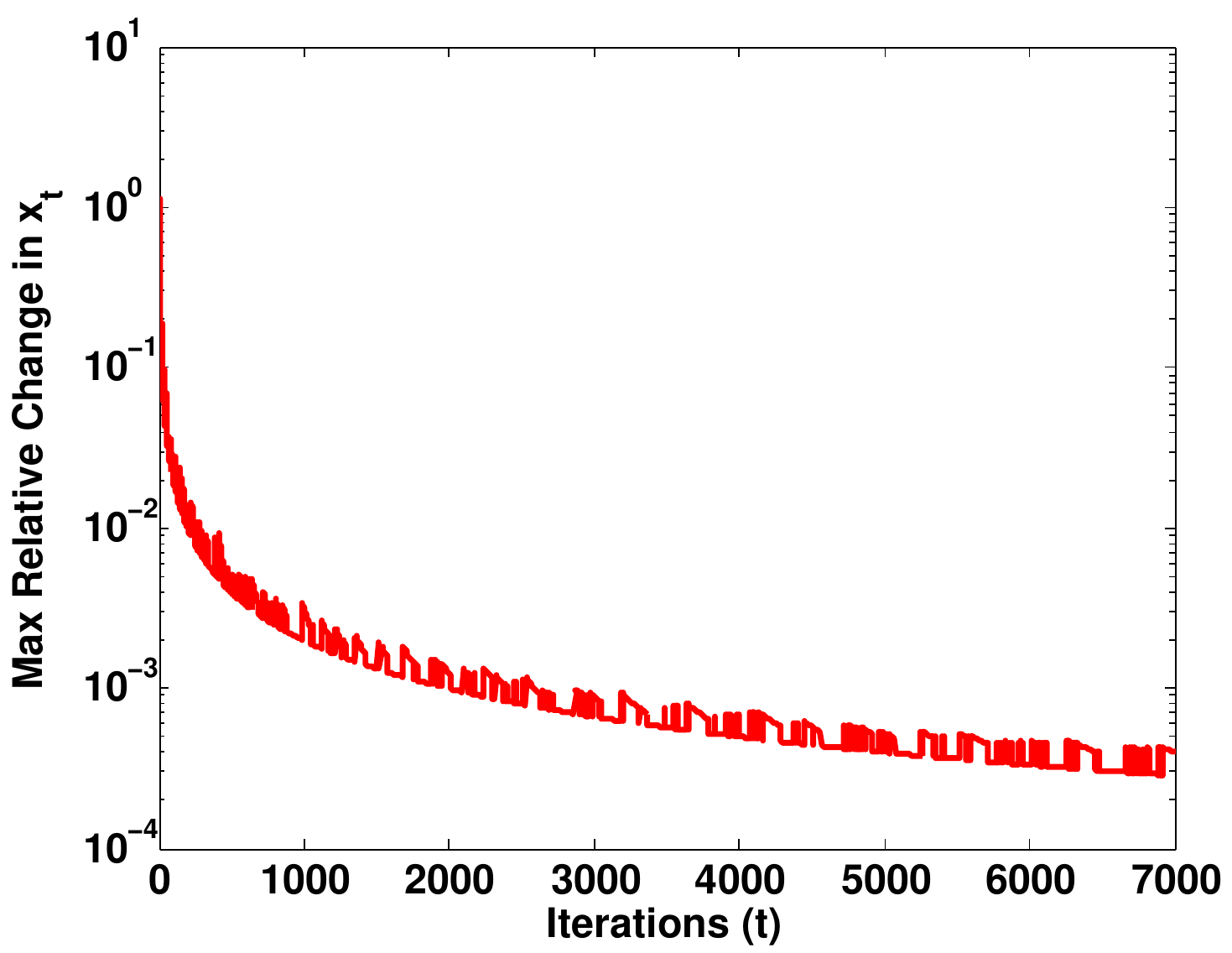}
\caption{Maximum relative change in the flow at any iteration with iterations. It can been seen that algorithm converges in terms of maximum relative change.}\label{fig:numres2}
\end{center}
\end{figure}

\begin{figure}[ht!]
\begin{center}
\includegraphics[width=0.6\linewidth]{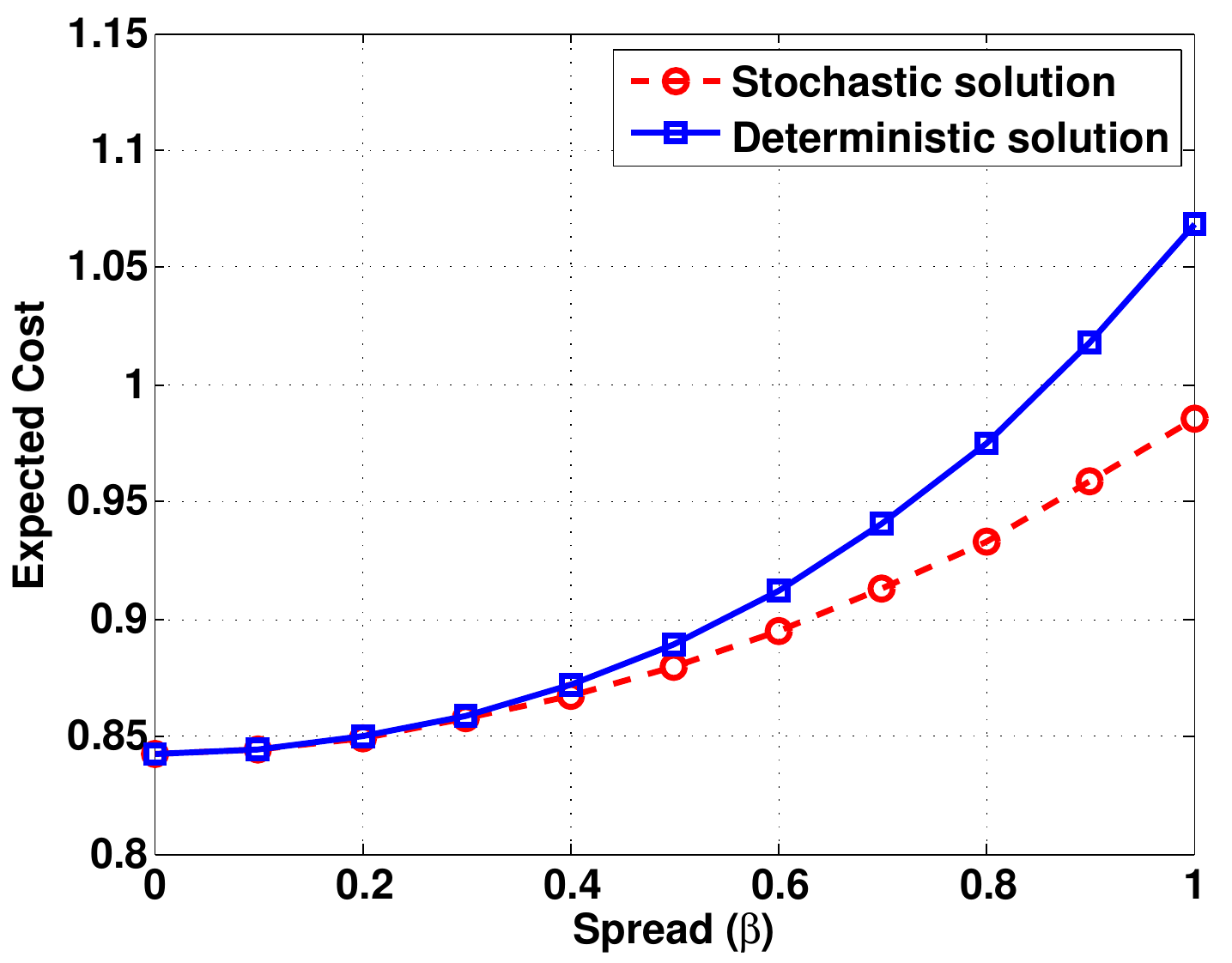}
\caption{Impact of randomness in the user flow.  The gain achieved by the proposed algorithm in  reducing the expected cost of the network is more evident in highly stochastic environment.}\label{fig:numres3}
\end{center}
\end{figure}

\textbf{Optimal Solution:}
The classic traffic assignment \cite{LEBLANC1975309} without considering the uncertainty in the system results in the optimal deterministic flow given by:

$$\mathbf{x} = \begin{bmatrix}
0.5238\\
0.5238\\
0.4762\\
0.4762
\end{bmatrix}$$
which may also be found by the solution $\alpha$ in the following problem:
\begin{align*}
\mathcal{P}: \alpha=  \arg\min_{0\le\alpha\le 1}&\left[ 2\alpha (0.3 + 0.6 \alpha ^4) \right.\\ &\left.+ 2(1- \alpha)(0.5 + 0.1(1-\alpha)^4 \right].
\end{align*}
Here, $\alpha$ denotes the units of traffic moving in  the upper path comprised of links 1 and 2 and (1-$\alpha$) units of traffic denotes the flow moving in the path comprised of links 3 and 4. In the stochastic environment ($\beta=1$), the optimum flow strategy to which gives the least travel costs on average turns out to be:
$$\mathbf{x} = \begin{bmatrix}
0.4206\\
0.4206\\
0.5794\\
0.5794
\end{bmatrix}$$
\abhiadd{which is very different than the solution in the deterministic environment. It shows that the stochastic nature of the flow affects the optimal flow significantly.}

\textbf{Comparison of Classical and Stochastic algorithm:} Fig. \ref{fig:numres1} shows the running mean of the total travel cost of the system for the classical deterministic  solution and the proposed stochastic solution in a stochastic environment with $\beta=1$. It is evident that the proposed stochastic solution performs much better in a random environment.

\textbf{Covergence of the proposed algorithm SFWTA: }
Fig. \ref{fig:numres2} shows the maximum relative change in elements of flow vector ({\em i.e. }$\mathbf{\detflow}$) as iterations progress for the stochastic environment with $\beta=1$. It can be seen that the maximum relative change decreases as iterations progress implying convergence of the proposed algorithm in the stochastic setting. Although there are sharp increases at some iterations, the overall trend is decreasing.

\textbf{Impact of randomness in the user flow:}
Fig. \ref{fig:numres3} shows the variation of cost with respect to spread parameter $\beta$. Recall that higher values of $\beta$ represents higher randomness in the user flow. As $\beta$ increases, the cost of the network increases and the difference between the stochastic and deterministic solution is more evident.

\section{Conclusion and Future Work}\label{conclusion}In this paper, we developed a framework for determining social optimally flow in  a stochastic environment and proposed an online stochastic Frank-Wolfe algorithm to compute optimal flow. It is motivated that there exists an element of randomness in the flow of every link in the network. The convergence of the algorithm is derived based on the proof developed in \cite{mokhtari2018stochastic}. 
We considered a simple example with the uniform uncertainty in the flow  and a socially optimal strategy in a stochastic environment was found using the proposed algorithm. The simulation results showed that the maximum relative change in
the elements of the deterministic-flow vector decreased as the iterations progressed implying convergence under the assumption of mutual independence of the random flows over different time instants. The solution obtained from the proposed model clearly performed better than the classical solution on average in a stochastic setting. Future works in this direction may include the study of determining optimum flows with stochastic network parameters which correspond to the realistic scenario of temporary road blockages and the effect of non-zero correlation between the random variables  and the evaluation of computational time-complexity for various algorithms.


\begin{thebibliography}{10}
\providecommand{\url}[1]{#1}
\csname url@samestyle\endcsname
\providecommand{\newblock}{\relax}
\providecommand{\bibinfo}[2]{#2}
\providecommand{\BIBentrySTDinterwordspacing}{\spaceskip=0pt\relax}
\providecommand{\BIBentryALTinterwordstretchfactor}{4}
\providecommand{\BIBentryALTinterwordspacing}{\spaceskip=\fontdimen2\font plus
\BIBentryALTinterwordstretchfactor\fontdimen3\font minus
  \fontdimen4\font\relax}
\providecommand{\BIBforeignlanguage}[2]{{%
\expandafter\ifx\csname l@#1\endcsname\relax
\typeout{** WARNING: IEEEtran.bst: No hyphenation pattern has been}%
\typeout{** loaded for the language `#1'. Using the pattern for}%
\typeout{** the default language instead.}%
\else
\language=\csname l@#1\endcsname
\fi
#2}}
\providecommand{\BIBdecl}{\relax}
\BIBdecl

\bibitem{wardrop1952some}
J.~G. Wardrop, ``Some theoretical aspects of road traffic research,'' in
  \emph{Proc. Inst Civil Engineers}, 1952.

\bibitem{LEBLANC1975309}
L.~J. LeBlanc, E.~K. Morlok, and W.~P. Pierskalla, ``An efficient approach to
  solving the road network equilibrium traffic assignment problem,''
  \emph{Transportation Research}, vol.~9, no.~5, pp. 309 -- 318, 1975.

\bibitem{ksaw_lit_rev}
K.~Saw, B.~K. Katti, and G.~Joshi, ``Literature review of traffic assignment:
  Static and dynamic,'' \emph{International Journal of Transportation
  Engineering}, vol.~2, no.~4, pp. 339--347, 2015.

\bibitem{Liu_2005}
W.~Y. Szeto and H.~K. Lo, ``Properties of dynamic traffic assignment with
  physical queues,'' \emph{Journal of the Eastern Asia Society for
  Transportation Studies}, vol.~6, pp. 2108--2123, 2005.

\bibitem{Michel_2013}
M.~Zou, X.~M. Chen, H.~Yu, Y.~Tong, Z.~Huang, M.~Li, and H.~Zou, ``dynamic
  transportation planning and operations: concept, framework and applications
  in china,'' \emph{Procedia-Social and Behavioral Sciences}, vol.~96, pp.
  2332--2343, 2013.

\bibitem{Han_2010}
S.~Han, S.~Fang, X.~Wang, X.~Chen, and Y.~Cai, ``A simulation-based dynamic
  traffic assignment model for emergency management on the hangzhou bay
  bridge,'' in \emph{Proc. ICCTP : Integrated Transportation Systems: Green,
  Intelligent, Reliable}, 2010, pp. 883--895.

\bibitem{mokhtari2018stochastic}
A.~Mokhtari, H.~Hassani, and A.~Karbasi, ``Stochastic conditional gradient
  methods: From convex minimization to submodular maximization,'' \emph{arXiv
  preprint arXiv:1804.09554}, 2018.

\bibitem{robb}
H.~Robbins and S.~Monro, ``A stochastic approximation method,'' \emph{Annals of
  Mathematical Statistics}, vol.~22, pp. 400--407, 1951.

\bibitem{hazan2016introduction}
E.~Hazan \emph{et~al.}, ``Introduction to online convex optimization,''
  \emph{Foundations and Trends{\textregistered} in Optimization}, vol.~2, no.
  3-4, pp. 157--325, 2016.

\bibitem{Rib_erg}
A.~Ribeiro, ``Ergodic stochastic optimization algorithms for wireless
  communication and networking,'' \emph{IEEE Transactions on Signal
  Processing}, vol.~58, no.~12, pp. 6369--6386, 2010.

\bibitem{M_trad_regret}
M.~Mahdavi, R.~Jin, and T.~Yang, ``Trading regret for efficiency: online convex
  optimization with long term constraints,'' \emph{Journal of Machine Learning
  Research}, vol.~13, no. Sep, pp. 2503--2528, 2012.

\bibitem{horn1990analog}
R.~A. Horn and R.~Mathias, ``An analog of the cauchy--schwarz inequality for
  hadamard products and unitarily invariant norms,'' \emph{SIAM Journal on
  Matrix Analysis and Applications}, vol.~11, no.~4, pp. 481--498, 1990.

\end{thebibliography}

\end{document}